\documentclass[aps,pra,twocolumn,a4paper,showpacs]{revtex4}
\usepackage{amsmath}
\usepackage{amssymb}
\usepackage{color}
\usepackage{epsfig}
\usepackage{bm}

\begin{document}

\title{Deviations of the exciton level spectrum in Cu$_2$O from the hydrogen 
series}
\author{F. Sch\"one}
\author{S.-O. Kr\"uger}
\author{P. Gr\"unwald}
\author{H. Stolz}
\author{S. Scheel}
\affiliation{Institut f\"ur Physik, Universit\"at Rostock,
Albert-Einstein-Strasse 23, D-18059 Rostock, Germany}
\author{M. A{\ss}mann}
\author{J. Heck\"otter}
\author{J. Thewes}
\author{D. Fr\"ohlich}
\author{M. Bayer}
\affiliation{Experimentelle Physik 2, Technische Universit\"at Dortmund, 
D-44221 Dortmund, Germany}

\begin{abstract}
Recent high-resolution absorption spectroscopy on excited excitons in cuprous 
oxide [Nature \textbf{514}, 343 (2014)] has revealed significant 
deviations of their spectrum from that of the ideal hydrogen-like series. 
Here we show that the complex band dispersion of the crystal, determining the 
kinetic energies of electrons and holes, strongly affects the exciton binding 
energies. Specifically, we show that the nonparabolicity of the band dispersion 
is the main cause of the deviation from the hydrogen series. Experimental 
data collected from high-resolution absorption spectroscopy in electric 
fields validate the assignment of the deviation to the nonparabolicity of 
the band dispersion.
\end{abstract}


\maketitle


\section{Introduction} Fluorescence spectra of atomic systems are an extremely 
rich source of information about the interactions of electrons and the nuclei, 
and their study laid the foundations of the development of quantum mechanics. 
The basic hydrogen dependence of the electron binding energies of $1/n^2$
already turns out to be insufficient for alkali, i.e. hydrogen-like, atoms 
where the polarisability of the ionic core modifies the Coulomb potential felt 
by the valence electron.
In semiconductor physics, the exciton concept translates the bound states of an 
electron-hole pair onto a hydrogen-like series, where the crystal environment 
is included in the effective masses and the dielectric function of the material 
\cite{Frenkel,Wannier,Mott}. However, the periodic arrangement of the crystal 
constituents breaks the rotational symmetry of atoms, and leads to 
semiconductor-specific effects such as detailed level splittings
\cite{Uihlein1979}. These splittings demonstrate deviations from the Rydberg 
formula.

Here we calculate the correct binding energies of $\mathrm{Cu_2O}$ from the 
band structure and cast them into the form of single-parameter corrections to 
the Rydberg series. In contrast to atoms, the origin of the excitonic 
correction is not a modification of the Coulomb potential, but the 
deviation from the parabolic band dispersion and hence the kinetic energy of 
electrons and holes. Based on the group-theoretical band Hamiltonian of Suzuki 
and Hensel (SH) \cite{Suzuki1974} adapted to the exact band structure of 
cuprous oxide derived using density functional theory \cite{French2009},we 
compute the exciton binding energies and extract from 
them the relevant corrections to the Rydberg series. Our theoretical results 
are compared to experimental data collected by high-resolution absorption 
spectroscopy. In combination with electric fields, we are able to extract the 
binding energies of excitons of $S$-, $P$-, $D$- and $F$-type, extending 
previous studies to far higher principal quantum numbers. The excellent 
agreement between theory and experiment corroborates our assignment of the 
deviations of the exciton level spectrum to the nonparabolicity of the band 
dispersion.


\section{Valence band dispersions in cuprous oxide}
Cuprous oxide ($\mathrm{Cu_2O}$) was historically the first material in which 
excitons were observed \cite{Gross52,Hayashi52,Gross56}. Their discovery, 
combined with the availability of exceptionally pure natural 
crystals \cite{Brandt2007}, sparked 
considerable interest in light-matter interactions in condensed-matter systems.
Cuprous oxide is endowed with a relatively large Rydberg energy of around 
$86$~meV. The crystal environment is taken into account by the permittivity 
$\varepsilon$ that, for crystals with cubic ($O_h$) symmetry such as 
$\mathrm{Cu_2O}$, is typically isotropic. This means that the exciton spectrum 
should simply be a scaled hydrogen spectrum. However, recent high-precision 
measurements of the $P$-exciton energies of the yellow series of 
$\mathrm{Cu_2O}$ showed a systematic deviation of the observed spectrum from 
the hydrogen analogy expectation \cite{Kazimierczuk2014}, in addition to the 
splitting of the excitons with different angular momenta for low 
$n$ \cite{Uihlein1979}. Several effects such as the influence of a frequency- and 
wavevector-dependent permittivity $\varepsilon(\mathbf{k},\omega)$, the 
coupling to LO-phonons as well as exchange interactions have all been proposed 
as causes for that deviation \cite{Kavoulakis1997}. In two-dimensional 
materials such as transition metal dichalgogenides (TMDC), the nonlocal nature 
of the Coulomb screening causes similar effects \cite{Chernikov14,Ye14,He14}. 
However, as we will show here, the deviation of the exciton binding energies of 
the yellow series in bulk cuprous oxide from the hydrogenic Rydberg series is a 
result of the nonparabolic dispersion of the highest valence band (with symmetry 
$\Gamma_7^+$ at the Brillouin zone center).


The description of excitons in a semiconductor requires the detailed knowledge 
of the electronic band structure. The common approach in solid-state physics is 
to approximate the extrema of valence and conduction bands via parabolic 
shapes. In the effective mass approximation these parabolas are then 
interpreted as the kinetic energy terms of electron and hole, respectively. A 
simple two-band model yields the Wannier equation, which is analogous to the
Schr\"odinger equation for hydrogen, and whose bound state solutions are the 
excitons \cite{HaugKoch} following the simple Rydberg formula.

Additional interactions such as spin-orbit and interband interactions lead to
further splitting and deformation of the relevant bands under consideration.
In $\mathrm{Cu_2O}$, the highest valence band with symmetry $\Gamma_5^+$ splits
under the spin-orbit interaction $\mathcal{H}_\mathrm{so}=\frac{1}{3} \Delta
\left(\mathbf{I}\cdot\bm{\sigma}\right)$, where $\mathbf{I}$ denotes the 
angular-momentum matrices for $I=1$ and $\bm{\sigma}$ the Pauli spin matrices,
into one (doubly degenerate) $\Gamma_7^+$- and two (doubly
degenerate) $\Gamma_8^+$-bands, associated with total angular momenta $J=1/2$
and $J=3/2$, respectively \cite{Suzuki1974}. The value of the spin-orbit 
splitting in $\mathrm{Cu_2O}$ is $\Delta=131$~meV. From symmetry arguments 
\cite{Luttinger1956} it follows that band interactions can be taken into 
account in a $6\times6$-matrix Hamiltonian that includes powers of the momentum 
$\mathbf{k}$ up to second order \cite{Suzuki1974}, 
\begin{gather}
\mathcal{H}_k = \frac{\hbar^2}{2m_e}\bigg( \left[A_1 +B_1 (\mathbf{I}\cdot
\bm{\sigma})\right]k^2
\nonumber \\
+\left[A_2 \left(I_x^2 -\frac{1}{3}I^2\right)
+ B_2\left(I_x\sigma_x - \frac{1}{3}\mathbf{I}\cdot
\bm{\sigma}\right) \right] k_x^2 + \mathrm{c.p.}
\nonumber \\
+\left[A_3(I_xI_y + I_yI_x)
+B_3(I_x\sigma_y + I_y\sigma_x)\right] \{k_xk_y\}
+ \mathrm{c.p.} \bigg),
\label{eq:SH}
\end{gather}
where $\mathrm{c.p.}$ stands for cycling permutations and
$\{k_xk_y\}=(k_xk_y+k_yk_x)/2$.
The band interaction Hamiltonian $\mathcal{H}_k$ contains a total of six
dimensionless parameters $\{A_{1\cdots3},B_{1\cdots3}\}$, which can be obtained
by matching the resulting energy dispersions to the exact band structure
derived from spin-DFT calculations \cite{French2009}. Note that DFT does not 
give the correct gap energy, which can be obtained experimentally.

\begin{figure}[htb]
\includegraphics[width=\columnwidth]{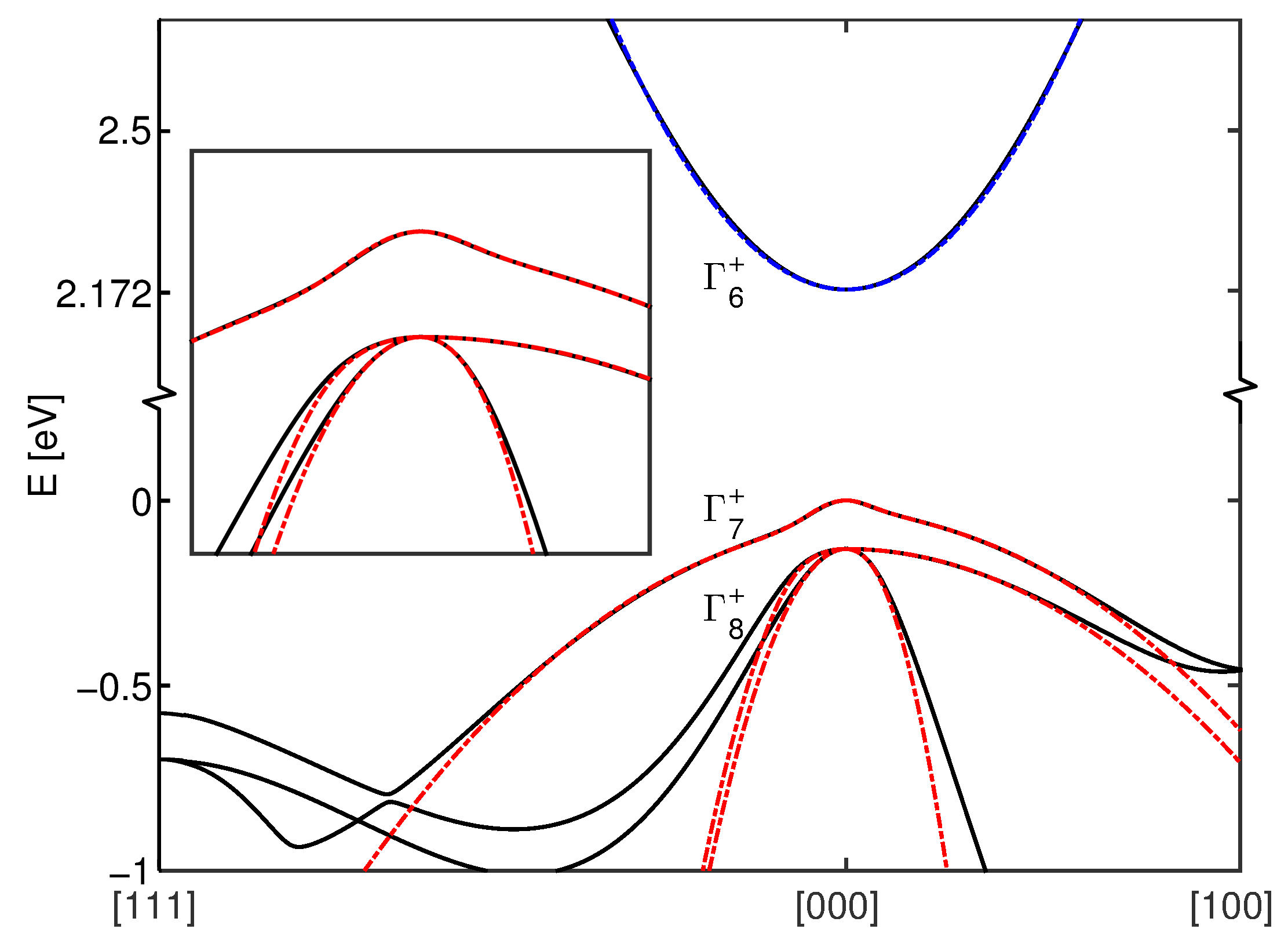}
\caption{Relevant band dispersions along two particular directions in the
Brillouin zone. Black lines: band dispersions from spin-DFT \cite{French2009}.
Blue dashed line: parabolic model for the lowest conduction ($\Gamma_6^+$) band
with an effective electron mass $m_e^* = 0.98\,m_e$. Red dashed lines:
best fits of the dimensionless parameters $\{A_{1\cdots3},B_{1\cdots3}\}$ in
$\mathcal{H}_k$ [Eq.~(\ref{eq:SH})] to  the band structure. The inset shows the 
uppermost valence bands and the corresponding fits in the vicinity of the 
$\Gamma$-point.}
\label{fig:02}
\end{figure}

The fit was performed for the highest valence band ($\Gamma_5^+$) which splits 
due to the spin-orbit interaction into an energetically higher $\Gamma_7^+$ and 
the two lower-energy $\Gamma_8^+$ bands. This yields three individual energy 
dispersions $E_{SH}^{(i)} (\mathbf{k})$ ($i=7,8hh,8lh$) for each direction in 
the Brillouin zone. 
A least-squares fit to the $\Gamma_7^+$ band that is of interest to us yields 
the parameters $A_1=-1.76$, $A_2=4.519$, $A_3=-2.201$, $B_1=0.02$, $B_2=-0.022$, 
and $B_3=-0.202$.
The lowest conduction band ($\Gamma_6^+$) can easily be
approximated by a parabola in the vicinity of the $\Gamma$-point and it has an 
isotropic curvature in all directions. The bound states formed between the 
$\Gamma_6^+$ and the $\Gamma_7^+$, $\Gamma_8^+$ bands are the excitons of the 
yellow and green series, respectively. Figure~\ref{fig:02} shows the relevant 
bands and the fits to the Hamiltonian
$\mathcal{H}_\mathrm{SH}=\mathcal{H}_\mathrm{so}+\mathcal{H}_k$. In the 
following, we neglect the anisotropy of the band dispersions and use a weighted 
average $\overline{E}_{SH}^{(i)}$ for simplicity, with the weights being the 
respective geometric multiplicities. After this procedure, there is no free 
parameter left. Furthermore, in doing so, the exciton angular momentum 
$l=0,1,2,\ldots$ becomes a good quantum number for labelling the states. Tiny 
fine structure splittings as reported recently are neglected \cite{Thewes15}.

\section{Exciton binding energies}
In the following we restrict our analysis to the excitons of the yellow series
($\Gamma_6^+\otimes\Gamma_7^+$) while an investigation of the excitons of the
green series ($\Gamma_6^+\otimes\Gamma_8^+$) would follow a similar path.
We first split the valence band dispersion $\overline{E}_{SH}^{(7)}$ into a 
parabolic part $T_h$ near $\mathbf{k}=\mathbf{0}$ and a nonparabolic 
contribution $\Delta T_h(k_h)$. Adding the kinetic energy $T_e$ of the 
conduction band as well as the Coulomb interaction $V_{e-h}$ to the valence band 
dispersion results in a Wannier equation that, due to the nonparabolic valence 
band dispersion, no longer resembles a simple Schr\"odinger-type equation. 
After transformation to the single-particle (exciton) picture and neglecting 
the center-of-mass momentum (i.e. setting $\mathbf{K}=\mathbf{0}$), the 
Hamiltonian takes the general form
\begin{equation}
\label{eq:modifiedH}
\mathcal{H}=\frac{\hbar^2k^2}{2\mu}+\Delta T_h+V_{e-h}(\mathbf{k})\star
\end{equation}
with the reduced exciton mass $\mu$ defined as usual. Here, the symbol $\star$ 
denotes the convolution operator.

The exciton binding energies $E_{n,l}$ are then obtained from the
eigenvalue equation $\mathcal{H}\Phi(\mathbf{k})=E_{n,l}\Phi(\mathbf{k})$
in momentum space which, due to the convolution with the Coulomb potential, is
in fact an integral equation. Without the nonparabolic contribution $\Delta
T_h$, the solution for the momentum-space wavefunction $\Phi(\mathbf{k})$ is
that for hydrogenic systems \cite{Podolsky1929}, and the energy eigenvalues
$E_{n,l}$ follow a hydrogen-like Rydberg series. Inclusion of the term
$\Delta T_h$ prevents one from obtaining an analytic solution.

In this case, we make use of a method proposed in Ref.~\cite{Szmytkowski2012}
by which the original eigenvalue problem
$\mathcal{H}\Phi(\mathbf{k})=E_{n,l}\Phi(\mathbf{k})$ is converted into an 
auxiliary Sturmian eigenvalue problem
\begin{equation}
\label{eq:sturmian}
\left( \frac{\hbar^2k^2}{2\mu}+\Delta T_h -E_{n,l} \right)
\Phi(\mathbf{k}) = -\lambda(E_{n,l}) V_{e-h}(\mathbf{k})\star\Phi(\mathbf{k})
\end{equation}
with Sturmian eigenvalue $\lambda(E_{n,l})$ that itself depends
parametrically on the sought energy eigenvalue $E_{n,l}$. The energy
eigenvalues of the system now emerge from the $\lambda$-spectrum if the
condition $\lambda(E_{n,l}) = 1$ is met. The advantage of this method 
is that Eq.~(\ref{eq:sturmian}) can be transformed, after separation of radial 
and angular variables, into a Fredholm integral equation of the second kind, 
whose real and symmetric integral kernel can be spectrally decomposed into 
eigenfunctions $g_{n,l}(k)$.

In the absence of any nonparabolicity ($\Delta T_h=0$), the eigenfunctions
$g_{n,l}(k)$ are orthonormal, and the condition $\lambda(E_{n,l})=1$
yields the standard hydrogenic Rydberg formula \cite{Szmytkowski2012}. The
inclusion of $\Delta T_h$ results in a mixing of eigenfunctions $g_{n,l}(k)$
with different principal quantum numbers $n$. Therefore, an additional
diagonalisation step has to be added. The numerical procedure thus involves:
\begin{enumerate}
\item solving the integral equation (\ref{eq:sturmian}) for all principal
quantum numbers $n$ with $n\leq n_{\max}$ for a given angular quantum number
$l$, with $E_{n,l}$ as parameters,
\item diagonalising the matrix of (non-orthogonal) eigenfunctions $g_{n,l}(k)$,
\item and extracting the energy eigenvalues that fulfil the condition
$\lambda(E_{n,l})=1$.
\end{enumerate}
Typically, principal quantum numbers up to $n_{\max}=60$ provided good
convergence of the numerical procedure to obtain the exciton spectrum up to
$n=25$.


\section{Deviation from the Rydberg series}

The energy eigenvalues $E_{n,l}$ that are obtained from such a
diagonalisation do no longer follow the Rydberg series of a hydrogen atom. 
In semiconductor physics, the common approach at this point is to employ 
$\mathbf k\cdot\mathbf p$-theory to account for the nonspherical 
symmetry \cite{Schechter,Baldereschi,Bernholc}. Incorporating the Luttinger 
parameters of $\mathrm{Cu_2O}$, it allows one to determine the binding energies 
as demonstrated in Ref.~\cite{Thewes15}. However, a few remarks are in order in
comparison to our approach. Besides the band gap and the Rydberg-energy,
applying $\mathbf k\cdot\mathbf p$-theory to cuprous oxide requires the
Luttinger parameters and in some cases the dielectric function, which are both
determined by fitting the model Hamiltonian to the experimental band structure.
In contrast, our approach fits the most general Hamiltonian for the cubic
crystal symmetry to an ab initio calculation of the band structure, thus
avoiding any experimental fit parameters at this point. Furthermore, the
solution of the coupled differential equations in the $\mathbf k\cdot\mathbf
p$-approach requires a set of basis functions that are arbitrary at this point.
On the other hand, the structure of the Fredholm integral equation yields a
natural basis set $g_{n,l}$. Another advantage of our method is
that we describe one source of deviations at one time. Combined with our purely
theoretical data we are thus able to gauge the influence of the nonparabolicity
on the deviation of exciton binding energies from the Rydberg series.

Phenomenologically, the energy shift of the excitonic Rydberg series can also
be traced back to deviations of the interaction potential between electron and
hole from the strict Coulomb potential at short distances (central-cell
corrections). For excitons in $\mathrm{Cu_2O}$ this was done for the
first time in Ref.~\cite{Washington1977}. There, the atomic 
concepts of quantum defects was applied, but only to match experimental data of
the yellow $S$-excitons, with the
assumption that the quantum defect for the $P$-excitons is negligibly small.
Besides the nonparabolicity of the bands, the binding energy of the excitons
can be affected by other central-cell corrections such as the coupling of
electrons and holes to LO-phonons, the frequency-dependent dielectric function
$\varepsilon(\mathbf{k},\omega)$ and exchange 
interactions \cite{Kavoulakis1997}.
However, as we are interested in excitons with large principal quantum number
$n$, their low binding energy prevents the coupling to LO-phonons. Possible 
coupling to phonons above the band gap is suppressed due the limitation of 
phonons by the low temperature. In addition, a large exciton Bohr radius allows 
one to assume that the dielectric function can be treated as a constant 
$\varepsilon=7.5$ \cite{Carabatos68}, hence we expect the contribution of the 
other central-cell corrections to be negligibly small. Overall, one may expect 
the nonparabolicity of the valence to be the major source of corrections to the 
ideal Rydberg series.

\begin{figure}[th]
\includegraphics[width=\columnwidth]{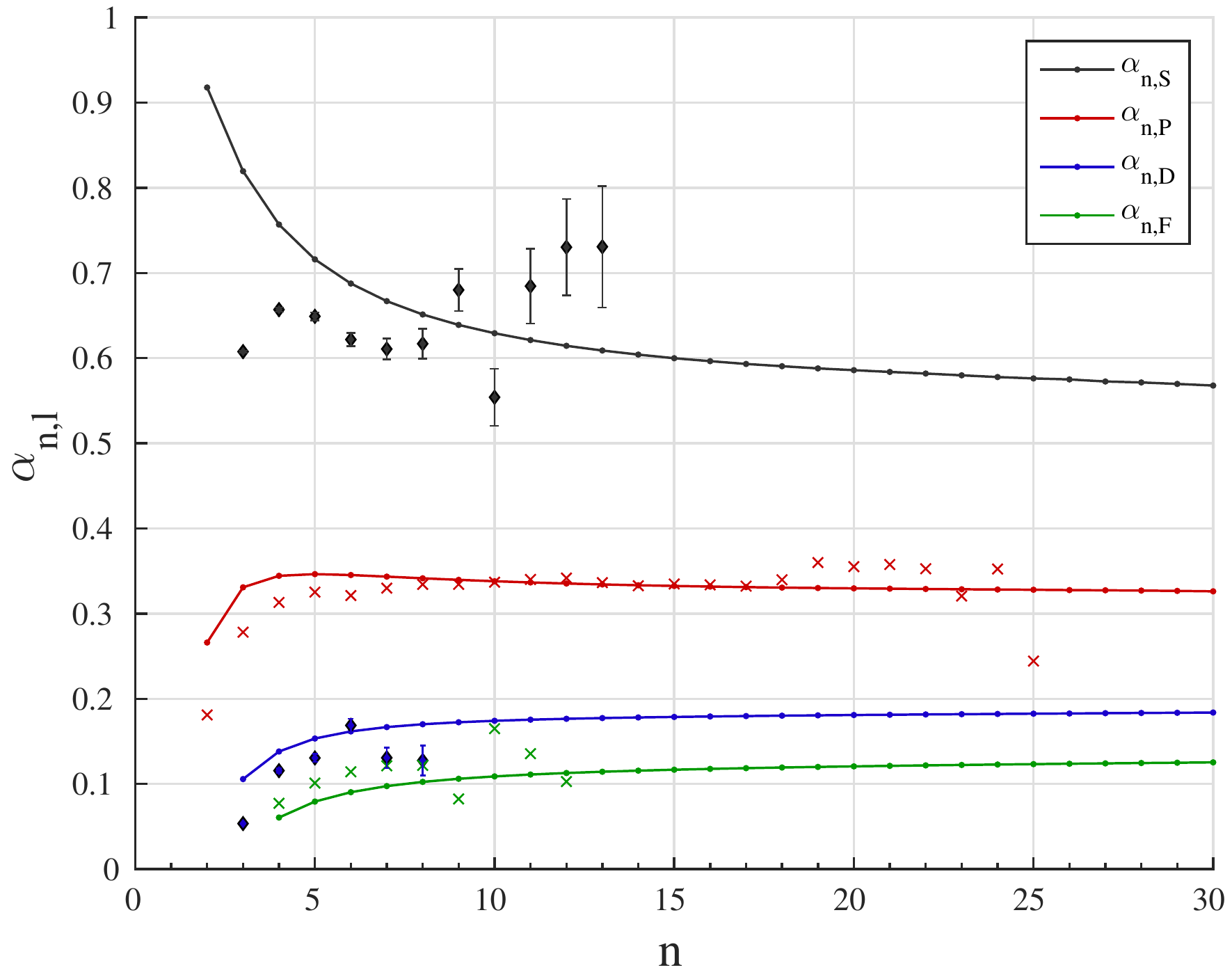}
\caption{\label{fig:qd} Deviations $\alpha_{n,l}$ from the Rydberg series
of the yellow excitons in
$\mathrm{Cu_2O}$ as function of principal quantum number $n$ for angular 
momenta $l=0\ldots3$. Solid lines: energy eigenvalues from 
Eq.~(\ref{eq:sturmian}). Marker: experimental values ($S$ -- black dots, $P$ -- 
red crosses, $D$ -- blue dots, $F$ -- green crosses) from absorption 
spectroscopy.}
\end{figure}

As the deviations from the ideal Rydberg series are small compared to the 
binding energies and tend to zero for large $n$, it is convenient to display 
these corrections relative to ideal variations of binding energies as
\begin{equation}
  \alpha_{n,l}=\left|\frac{E_n-E_n^*}{E_{n+1}^*-E_n^*}\right|,
\end{equation}
where $E_n$ is the actually measured binding energy (difference to band 
gap), while the $E_\nu^*$ represent ideal Rydberg energies $Ry/\nu^2$.
The results are shown in Fig.~\ref{fig:qd} (solid lines) as 
function of principal quantum number $n$ for the lowest angular momentum states 
with $l=0\ldots3$. 
As expected from the modified Hamiltonian (\ref{eq:modifiedH}), the theoretical 
corrections saturate for large $n$, and decrease with increasing $l$.

\section{Transmission spectroscopy} To determine the energies of exciton 
states in $\mathrm{Cu_2O}$ experimentally, we performed high resolution 
absorption spectroscopy on a high-quality crystal. The energies of the 
$P$-excitons can be accessed directly by one-photon absorption in 
electric-dipole approximation, as demonstrated in \cite{Kazimierczuk2014}. In a 
strict sense angular momentum is not a good quantum number in crystals, but can 
be used approximately in high-symmetry cubic crystals with $O_h$ or $T_d$ 
symmetry. The symmetry reduction leads to an admixture of $P$-excitons to the 
$F$-excitons so that they become observable in absorption \cite{Uihlein1979}. As 
a consequence, the energies of these states can be determined with high 
accuracy \cite{Thewes15}. In case they show a fine structure splitting, we have 
taken the center of gravity energy of the corresponding multiplet, which is in 
line with the theoretical model that averages over different $k$-space 
directions. While the energies of excitons with odd-parity envelopes can be 
determined accurately in that way, this is not possible for states with even 
parity.

\begin{figure}[ht]
\centerline{\includegraphics[width=\columnwidth]{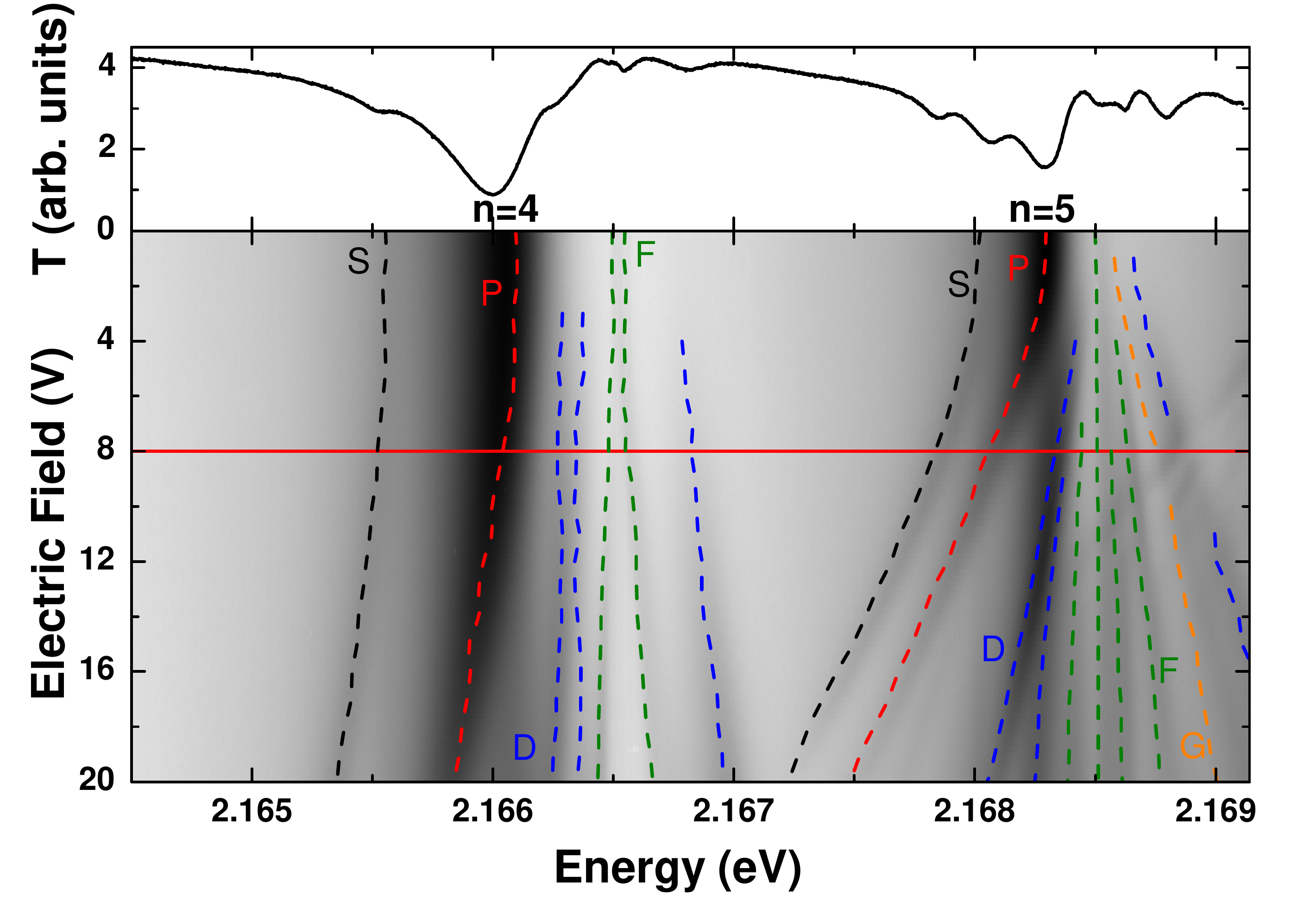}}
\caption{Absorption spectra of the $n=4$ and $n=5$ excitons in external 
electric fields. The curve on top shows a cut through the contour plot at 
$U=8$~V. }
\label{fig:Spectrum}
\end{figure}

Therefore, we applied additionally an electric field along the optical axis, 
which leads to a mixing of the $S$-exciton with the $P$-exciton with zero 
magnetic quantum number. Similarly, $D$-excitons become mixed with $P$- and 
$F$-excitons of the same magnetic quantum number. As a consequence, the related 
features appear in the absorption spectra, see Fig.~\ref{fig:Spectrum}, which 
shows the absolute absorption of the $n=4$ and $n=5$ excitons as function of the
electric field. With increasing field strength new features emerge and 
intensify, as expected from the increased state mixing with the $P$-excitons. 
Also line splittings are observed which ultimately lead to the Stark ladder. 
From extrapolation towards $U = 0$, the energies of these excitons at zero 
field can be accurately assessed. From their order in energy we can attribute 
them to $S$- and $P$-excitons, as labeled.

\begin{figure}[ht]
\centerline{\includegraphics[width=\columnwidth]{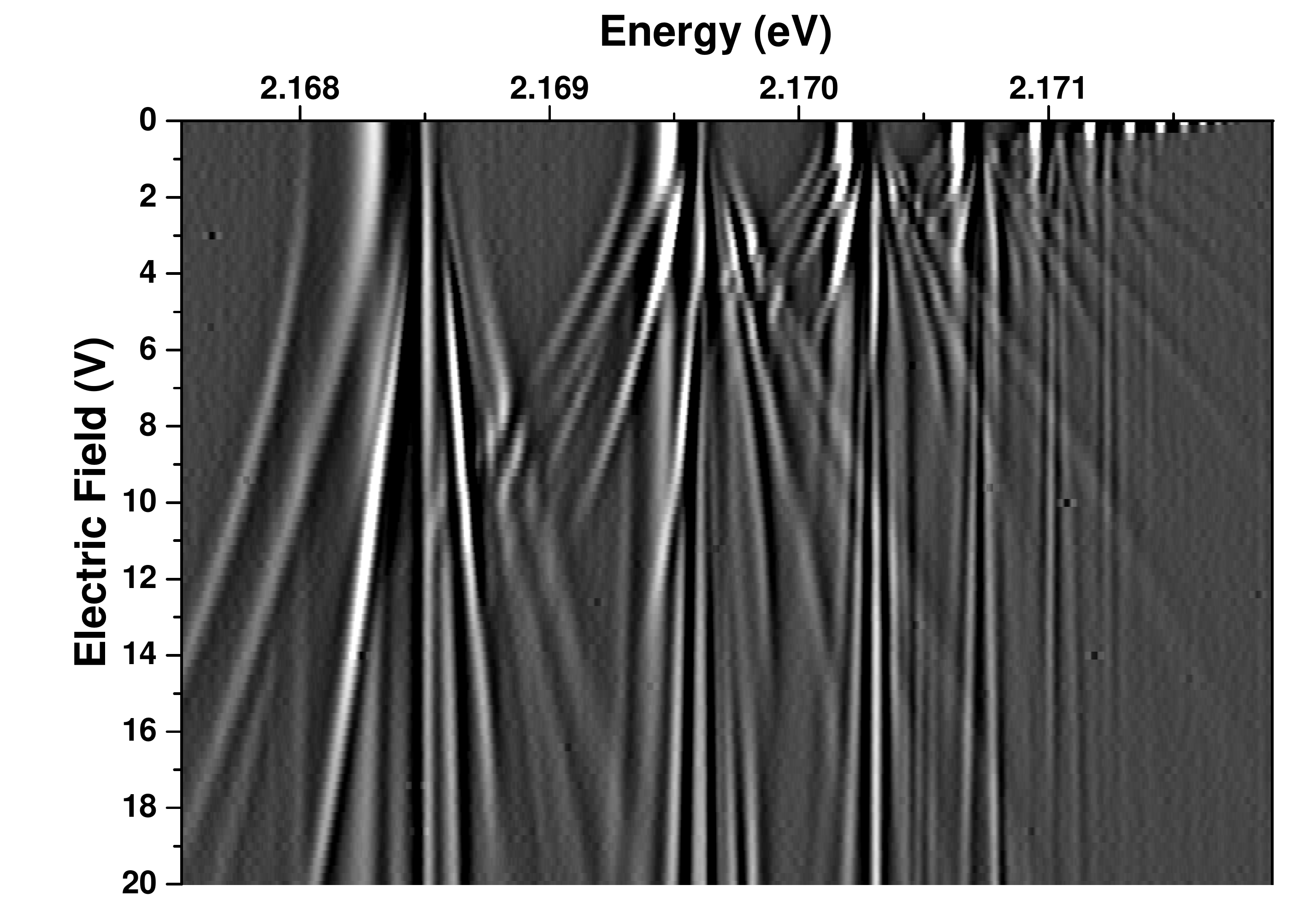}}
\caption{\label{fig:derivative} Contour plot of the second derivative of an 
absorption spectrum extended over a larger energy range covering the excitons 
from $n=5$ towards the band gap. The multiple splitting of the states while 
transforming into the Stark ladder becomes obvious.}
\end{figure}

To enhance the accuracy of the energy evaluation, we have also taken the second 
derivatives of the absorption spectra, shown in Fig.~\ref{fig:derivative}. By 
doing so, the enormous intensity differences in absolute transmission are 
leveled out, and the different absorption lines become much better resolvable, 
so that high-accuracy determination of the $S$- and $D$-exciton energies becomes
possible: For the $D$-excitons again the center-of-gravity is determined for 
each principal quantum number. In doing so, we are able to extend the reported 
energies to significantly higher principal quantum numbers, for the 
$S$-excitons up to $n=12$ compared to 7 in previous reports, and up to $n=8$ 
for the $D$-excitons compared to 5 previously. The assignment of the states 
via the second derivative is depicted in the transmission spectrum 
Fig.~\ref{fig:Spectrum} by the dotted lines.  Note that for higher $n$ these 
excitons become also optically activated in electric fields. However, it is no 
longer possible to assign them in the multiplicity of emerging lines, so that 
also their energies can no longer be determined with the required accuracy for 
assessing the deviation $\alpha_{n,l}$.

\section{Results and Discussion}
The experimentally extracted values for the deviations $\alpha_{n,l}$ of the 
exciton binding energies from the ideal Rydberg series are shown 
in Fig.~\ref{fig:qd}, together with the theoretical results for the energy 
eigenvalues that are obtained from solving Eq.~(\ref{eq:sturmian}). In 
addition, the $S$-excitons are affected by electron-hole exchange interaction 
with a common scaling of $n^{-3}$. The experimental binding energies of the 
$S$-excitons, whose deviation $\alpha_{n,S}$ is depicted in 
Fig.~\ref{fig:qd}, have thus been shifted downwards by an amount of 
$8.04~\mbox{meV}/n^3$ (following previous measurements of the exchange 
\cite{Uihlein1979}) with respect to the values extracted from the absorption 
spectra.

The agreement, in particular of the $P$-exciton resonances for large $n$, is 
astonishingly good. This implies that nonparabolicity of the bands (in case of 
$\mathrm{Cu_2O}$ only the valence bands) due to interband coupling is the 
dominant contribution to the deviation of the exciton resonances from the 
hydrogenic Rydberg series. Nonetheless, one still observes a systematic 
deviation of the $S$-exciton energies from their predicted values, in 
particular for low $n$.
This can be attributed to the fact that in this energy range the background 
permittivity $\varepsilon(\mathbf{k},\omega)$ is far from 
constant \cite{Dawson73}, and that the binding energy, especially for $n=2$, is 
close to a phonon resonance. Taking this into account will shift the data point 
closer to the theoretically expected value. This points to a possible extension 
of our theory in which a frequency-dependent permittivity could be incorporated 
in a self-consistent manner by including it both in the Coulomb interaction 
potential and the Rydberg energy.
The observed high-$n$ shift on the other hand may be due to the fact that 
the measurement setup had to be altered for the resonances beyond $n=10$. Thus,
a small systematic shift might have occurred, which features prominently in 
the deviation parameter $\alpha_{n,S}$.

Although we have focussed solely on the yellow exciton series of 
$\mathrm{Cu_2O}$, it is clear that our procedure is valid for all semiconductors 
that possess both spin-orbit interactions as well as cubic symmetry within 
their band structure. In particular, we expect the green exciton series of 
$\mathrm{Cu_2O}$ to provide purely negative values for $\alpha_{n,l}$
because the interband coupling results in an increased curvature of 
the $\Gamma_8^+$ valence band.

\acknowledgments We thank Ch. Uihlein for his suggestion to
analyse the data by use of the 2nd derivative. We gratefully acknowledge 
support by the Collaborative Research Centre SFB 652/3 'Strong correlations in 
the radiation field' and the International Collaborative Research Centre
TRR 160 'Coherent manipulation of interacting spin excitations in
tailored semiconductors', both funded by the Deutsche
Forschungsgemeinschaft.


\begin{thebibliography}{99}

\bibitem{Frenkel}
J.~Frenkel, \textit{On the transformation of light into heat in solids}, Phys. 
Rev. \textbf{37}, 1276 (1931).
\bibitem{Wannier}
G.H.~Wannier, \textit{The structure of electronic excitation levels in 
insulating crystals}, Phys. Rev. \textbf{52}, 191 (1937).
\bibitem{Mott}
N.F.~Mott, \textit{The conduction band and ultra-violet absorption of 
alkali-halide crystals}, Trans. Faraday Soc. \textbf{34}, 500 (1938).
\bibitem{Uihlein1979}
C.~Uihlein, D.~Fr\"ohlich, and R.~Kenklies, \textit{Investigation of exciton 
fine structure in $\mathrm{Cu_2O}$}, Phys. Rev. B \textbf{23}, 2731 (1981).
\bibitem{Suzuki1974}
K.~Suzuki and J.C.~Hensel, \textit{Quantum resonances in the valence bands of 
germanium}, Phys. Rev. B \textbf{9}, 4184 (1974).
\bibitem{French2009}
M.~French, R.~Schwartz, H.~Stolz, and R.~Redmer, \textit{Electronic band 
structure of $\mathrm{Cu_2O}$ by spin density functional theory}, J. Phys.: 
Condensed Matter \textbf{21}, 015502 (2009).
\bibitem{Gross52}
E.F.~Gross and N.A.~Karryjew, \textit{The optical spectrum of the exciton}, 
Dokl. Akad. Nauk SSSR \textbf{84}, 471 (1952).
\bibitem{Hayashi52}
M.~Hayashi and K.~Katsuki, \textit{Hydrogen-like absorption spectrum of 
cuprous oxide}, J. Phys. Soc. Japan \textbf{7}, 599 (1952).
\bibitem{Gross56}
E.F.~Gross, \textit{Optical spectrum of excitons in the crystal lattice}, Il 
Nuovo Cimento \textbf{4}, 672 (1956).
\bibitem{Brandt2007} J.~Brandt, D.~Fr\"ohlich, C.~Sandfort, M.~Bayer, H.~Stolz, 
and N.~Naka, \textit{Ultranarrow Optical Absorption and Two-Phonon Excitation 
Spectroscopy of $\mathrm{Cu_2O}$ Paraexcitons in a High Magnetic Field}, Phys. 
Rev. Lett. \textbf{99}, 217403 (2007).
\bibitem{Kazimierczuk2014}
T.~Kazimierczuk, D.~Fr\"ohlich, S.~Scheel, H.~Stolz, and M.~Bayer, 
\textit{Giant Rydberg excitons in the copper oxide $\mathrm{Cu_2O}$}, Nature
\textbf{514}, 343 (2014).
\bibitem{Kavoulakis1997}
G.M.~Kavoulakis, Y.-C.~Chang, and G.~Baym, \textit{Fine structure of excitons 
in $\mathrm{Cu_2O}$}, Phys. Rev. B \textbf{55}, 7593 (1997).
\bibitem{Chernikov14}
A.~Chernikov \textit{et al.},
\textit{Exciton Binding Energy and Nonhydrogenic 
Rydberg Series in Monolayer $\mathrm{WS_2}$}, Phys. Rev. Lett. \textbf{113}, 
076802 (2014).
\bibitem{Ye14}
Z.~Ye \textit{et al.},
\textit{Probing excitonic dark states in single-layer tungsten disulphide}, 
Nature \textbf{513}, 214 (2014).
\bibitem{He14}
K.~He \textit{et al.},
\textit{Tightly Bound Excitons in Monolayer $\mathrm{WSe_2}$}, 
Phys. Rev. Lett. \textbf{113}, 026803 (2014).
\bibitem{HaugKoch}
H.~Haug and S.W.~Koch, \textit{Quantum Theory of the Optical and Electronic
Properties of Semiconductors} (World Scientific, Singapore, 2009).
\bibitem{Luttinger1956}
J.M.~Luttinger, \textit{Quantum theory of cyclotron resonance in 
semiconductors}, Phys. Rev. \textbf{102}, 1030 (1956).
\bibitem{Thewes15}
J.~Thewes \textit{et al.},
\textit{Observation of High Angular Momentum Excitons in Cuprous Oxide}, Phys. 
Rev. Lett. \textbf{115}, 027402 (2015).
\bibitem{Podolsky1929}
B.~Podolsky and L.~Pauling, \textit{The momentum distribution in 
hydrogen-like atoms}, Phys. Rev. \textbf{34}, 109 (1929).
\bibitem{Szmytkowski2012}
R.~Szmytkowski, \textit{Alternative approach to the solution of the 
momentum-space Schr\"odinger equation for bound states of the N-dimensional 
Coulomb problem}, Ann. Phys. (Berlin) \textbf{524}, 345 (2012).
\bibitem{Schechter}
D. Schechter, \textit{Theory of shallow acceptor states in Si and Ge}, J. Phys. 
Chem. Solids \textbf{23}, 237 (1962). 
\bibitem{Baldereschi}
A.~Baldereschi and N.~O.~Lipari, \textit{Spherical Model of Shallow Acceptor 
States in Semiconductors}, Phys. Rev. B \textbf{8}, 2697 (1973).
\bibitem{Bernholc}
J.~Bernholc and S.~T.~Pantelides, \textit{Theory of binding energies of 
acceptors in semiconductors}, Phys. Rev. B \textbf{15}, 4935 (1977).
\bibitem{Washington1977}
M.A.~Washington \textit{et al.},
\textit{Spectroscopy of excited yellow exciton states in $\mathrm{Cu_2O}$ by 
forbidden resonant Raman scattering}, Phys. Rev. B \textbf{15}, 2145 (1977).
\bibitem{Carabatos68}
C.~Carabatos, A.~Diffin\'e, and M.~Sieskind, \textit{Contribution \`a l'\'etude 
des bandes fondamentales de vibration du r\'eseau de la cuprite (Cu2O)}, J. 
Physique \textbf{29}, 529 (1968).
\bibitem{Dawson73}
P.~Dawson, M.M.~Hargreave, and G.R.~Wilkinson, \textit{The dielectric and 
lattice vibrational spectrum of cuprous oxide}, J. Phys. Chem. Solids
\textbf{34}, 2201 (1973).
\end{thebibliography}
\end{document}